\documentclass[twocolumn,showpacs,preprintnumbers,amsmath,amssymb,superscriptaddress,prl]{revtex4}

\usepackage{graphicx}
\usepackage{times}
\usepackage{dcolumn}
\usepackage{bm}
\usepackage[pdftex,colorlinks,bookmarks=false,citecolor=blue,linkcolor=red,urlcolor=blue]{hyperref}
\usepackage{color}

\newcommand{\Sp}{\mathbf{S}}

\newcommand{\irrma}{Institut Romand de Recherche Num\'erique en Physique des Mat\'eriaux (IRRMA), CH-1015 Lausanne, Switzerland}

\begin{document}

\title{
Emergent multipolar spin correlations in a fluctuating spiral --\\
The frustrated ferromagnetic $S=1/2$ Heisenberg chain in a magnetic field}

\author{Julien Sudan}
\affiliation{%
\irrma
}%
\author{Andreas L\"uscher}
\affiliation{%
\irrma
}%
\author{Andreas M. L\"auchli}
\email{laeuchli@comp-phys.org}
\affiliation{Max Planck Institut f\"ur Physik komplexer Systeme, N\"othnitzer Str. 38, D-01187 Dresden, Germany}

\date{\today}

\begin{abstract}
We present the phase diagram of the frustrated ferromagnetic $S=1/2$ Heisenberg $J_1-J_2$ chain in a magnetic field,
obtained by large scale exact diagonalizations and density matrix renormalization group simulations.
A vector chirally ordered state, metamagnetic behavior and a sequence of spin-multipolar Luttinger liquid phases up 
to hexadecupolar kind are found. 
We provide numerical evidence for a locking mechanism, which can drive spiral states towards 
spin-multipolar phases, such as quadrupolar or octupolar phases. Our results also shed light on 
previously discovered spin-multipolar phases in two-dimensional $S=1/2$ quantum magnets in a magnetic field.
\end{abstract}

\maketitle
\paragraph{Introduction}
Spiral or helical ground states are an old and well understood concept in classical magnetism~\cite{OldSpirals}, 
and several materials are successfully described based on theories of spiral states.
For low spin and dimensionality however quantum fluctuations become important and might destabilize the spiral states. 
Given that spiral states generally arise due to competing interactions, fluctuations are expected to be particularly strong.

A prominent instability of spiral states is their intrinsic twist $\langle\Sp_i\times\Sp_j\rangle$ (vector chirality)~\cite{Villain}. 
It has been recognized that finite temperature~\cite{Kawamura} or quantum~\cite{Chandra91} fluctuations can disorder the 
spin moment $\langle\Sp_i\rangle$ of the spiral, while the twist remains finite. Such a state is called $p-$type spin 
nematic~\cite{Andreev84}.  
In the context of quantum fluctuations such a scenario has been confirmed recently in a ring-exchange model~\cite{Laeuchli05}, while
possible experimental evidence for the thermal scenario has been presented in \cite{Cinti08}.
The twist also gained attention in multiferroics, since it couples directly to the ferroelectricity~\cite{Seki08}.

In this Letter we provide evidence for the existence of yet a different instability of spiral states {\em towards spin-multipolar phases}.
The basic idea is that many spin-multipolar order parameters are finite 
 in the magnetically ordered spiral state, but that under a suitable
amount of fluctuations the primary spin order is lost, while a spin-multipolar order parameter survives.
We demonstrate this mechanism based on the magnetic field phase diagram of a prototypical model,
the frustrated $S=1/2$ Heisenberg chain with ferromagnetic nearest-neighbor and antiferromagnetic next nearest-neighbor
interactions. Furthermore we show that this instability provides a natural and unified understanding of previously discovered two-dimensional 
spin-multipolar phases~\cite{shannon06,momoi06}.
\begin{figure}[hbt]
\centering
\includegraphics[width=\linewidth]{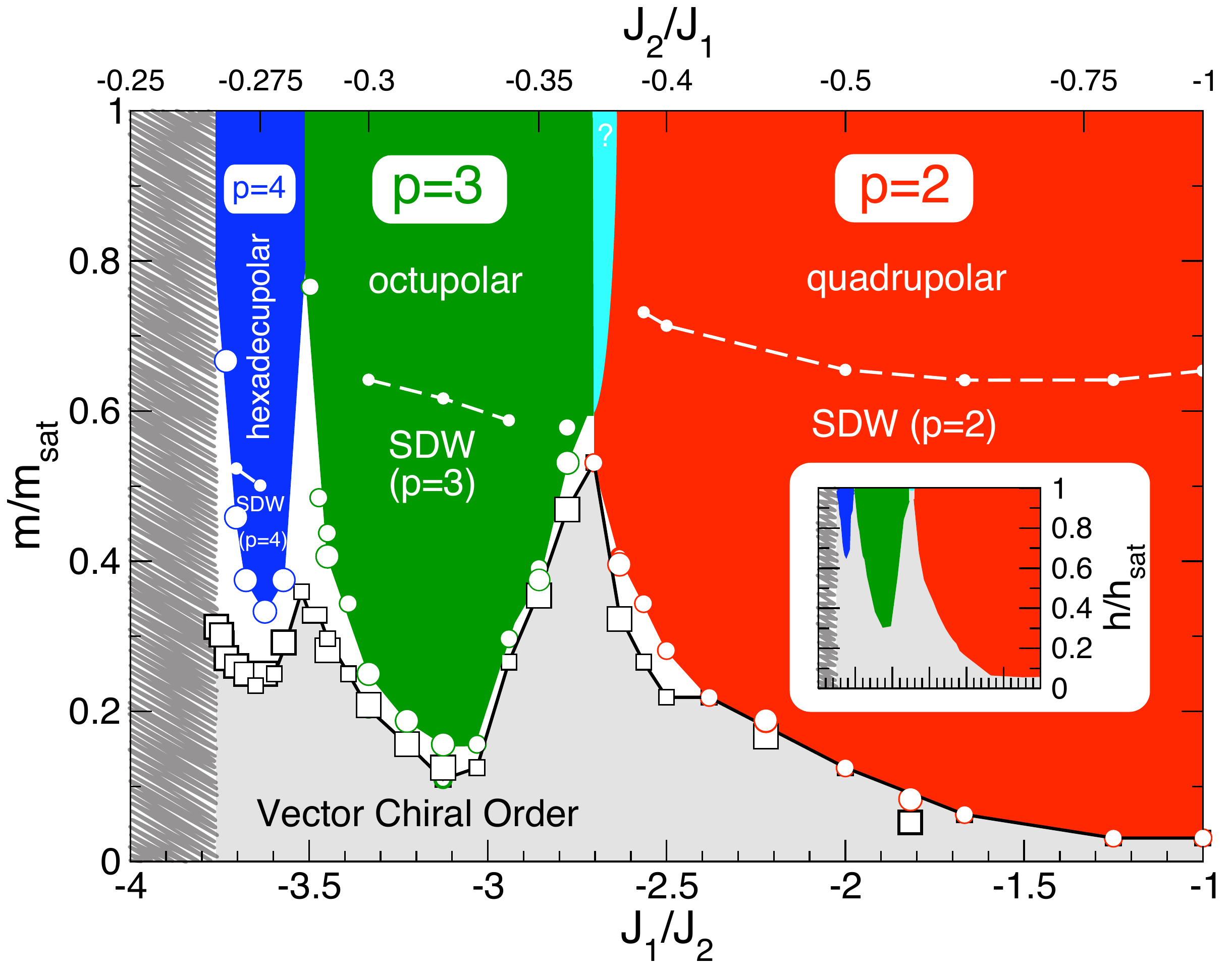}
\caption{(Color online)
Phase diagram of the frustrated ferromagnetic chain (\protect{\ref{eqn:Hamiltonian}}) in the $J_1/J_2$ vs. $m/m_\mathrm{sat}$ plane.
The grey low-$m$ region exhibits vector chiral long range order. The colored regions denote spin-multipolar Luttinger liquids of bound states of $p=2,3,4$ spin flips. Close to saturation the dominant correlations are multipolar, while below the dashed crossover lines, the dominant correlations are of SDW$(p)$-type. 
The tiny cyan colored region corresponds to an incommensurate $p=2$ phase.
The white region denotes a metamagnetic jump. Finally the scribbled region close
to the transition $J_1/J_2\rightarrow -4$ has not been studied here, but consists most likely of a low field vector chiral phase, followed by a 
metamagnetic region extending up to saturation magnetization.
The inset shows the same diagram in the $J_1/J_2$ vs. $h/h_\mathrm{sat}$ plane. 
}
\label{fig:PhaseDiagram}
\end{figure}
\paragraph{Hamiltonian and Methods}
To be specific, we determine numerically the phase diagram of the following Hamiltonian
\begin{equation}
H=J_1\sum_i\mathbf{S}_i\cdot\mathbf{S}_{i+1}+J_2\sum_i\mathbf{S}_i\cdot\mathbf{S}_{i+2}-h\sum_iS^z_i,
\label{eqn:Hamiltonian}
\end{equation}
and we set $J_1=-1$, $J_2\geq 0$ in the following. $\mathbf{S}_i$ are spin-$1/2$ operators at site $i$, while $h$ 
denotes the uniform magnetic field. 
The magnetization is defined as $m:=1/L\sum_i S^z_i$.
We employ exact diagonalizations (ED) on systems sizes up to $L=64$ sites complemented by density matrix renormalization group 
(DMRG)~\cite{White92} simulations on open systems of maximal length $L=384$, retaining up to $800$ basis states.

\paragraph{Overall Phase Diagram}
The classical ground state of the Hamiltonian~(\ref{eqn:Hamiltonian}) is ferromagnetic for $J_2< 1/4$ 
and a spiral with pitch angle $\varphi=\arccos(1/4J_2) \in [0,\pi/2]$ otherwise. The Lifshitz point is located at $J_2=1/4$. 
In a magnetic field the spins develop a uniform component along the field, while the pitch angle in the plane transverse 
to the field axis is unaltered by the field.

The zero field quantum mechanical phase diagram for $S=1/2$ is still unsettled. Field theoretical work predicts a finite, but tiny
gap accompanied by dimerization~\cite{Cabra00,ItoiQin01} for $J_2> 1/4$, which present numerical approaches are unable to resolve.
The classical Lifshitz point $J_2=1/4$ is not renormalized for $S=1/2$, and the transition point manifests itself on finite systems as a level crossing between the ferromagnetic multiplet and an exactly known singlet state~\cite{Hamada88}. 
The theoretical phase diagram at finite field has recently received considerable attention~\cite{Heidrichmeisner06,Kecke07,Vekua07}, triggered by 
experiments on quasi one-dimensional cuprate helimagnets~\cite{Hase04,Enderle05,Drechsler07}. One of the most peculiar features of the finite size magnetization
process is the appearance of elementary magnetization steps of $\Delta S^z=2,3,4$ in certain $J_2$ and $m$ regions. This has 
been attributed to the formation of bound states of spin flips, leading to dominant spin-multipolar correlations close to saturation. A detailed phase diagram is however still lacking.

We present our numerical phase diagram in the $J_2/|J_1|$ vs.~$m/m_\mathrm{sat}$ plane in Fig.~\ref{fig:PhaseDiagram}.
At least five different phases are present.
The low magnetization region consists of a single vector chiral phase (grey). Below the saturation magnetization
we confirm the presence of three different multipolar Luttinger liquid phases (red, green and blue). The red phase extends up to $J_2 \rightarrow \infty$~\cite{Vekua07},
and its lower border approaches $m=0^+$ in that limit. All three multipolar liquids present a crossover as a function of $m/m_\mathrm{sat}$, where the dominant 
correlations change from spin-multipolar close to saturation to spin density wave character at lower magnetization. One also expects 
a tiny incommensurate $p=2$ phase close to the $p=3$ phase~\cite{Kecke07}, which we did not aim to localize in this study.
Finally the multipolar Luttinger liquids are separated from the vector chiral phase by a
metamagnetic transition, which occupies a larger and larger fraction of $m$ as $J_2\rightarrow 1/4^+$, leading to an absence of 
multipolar liquids composed of five or more spin flips. 
In the following we will characterize these phases in more detail and put forward an explanation for the
occurrence and locations of the spin-multipolar phases.

\paragraph{Vector chiral phase}
\begin{figure}
\centering
\includegraphics[width=\linewidth]{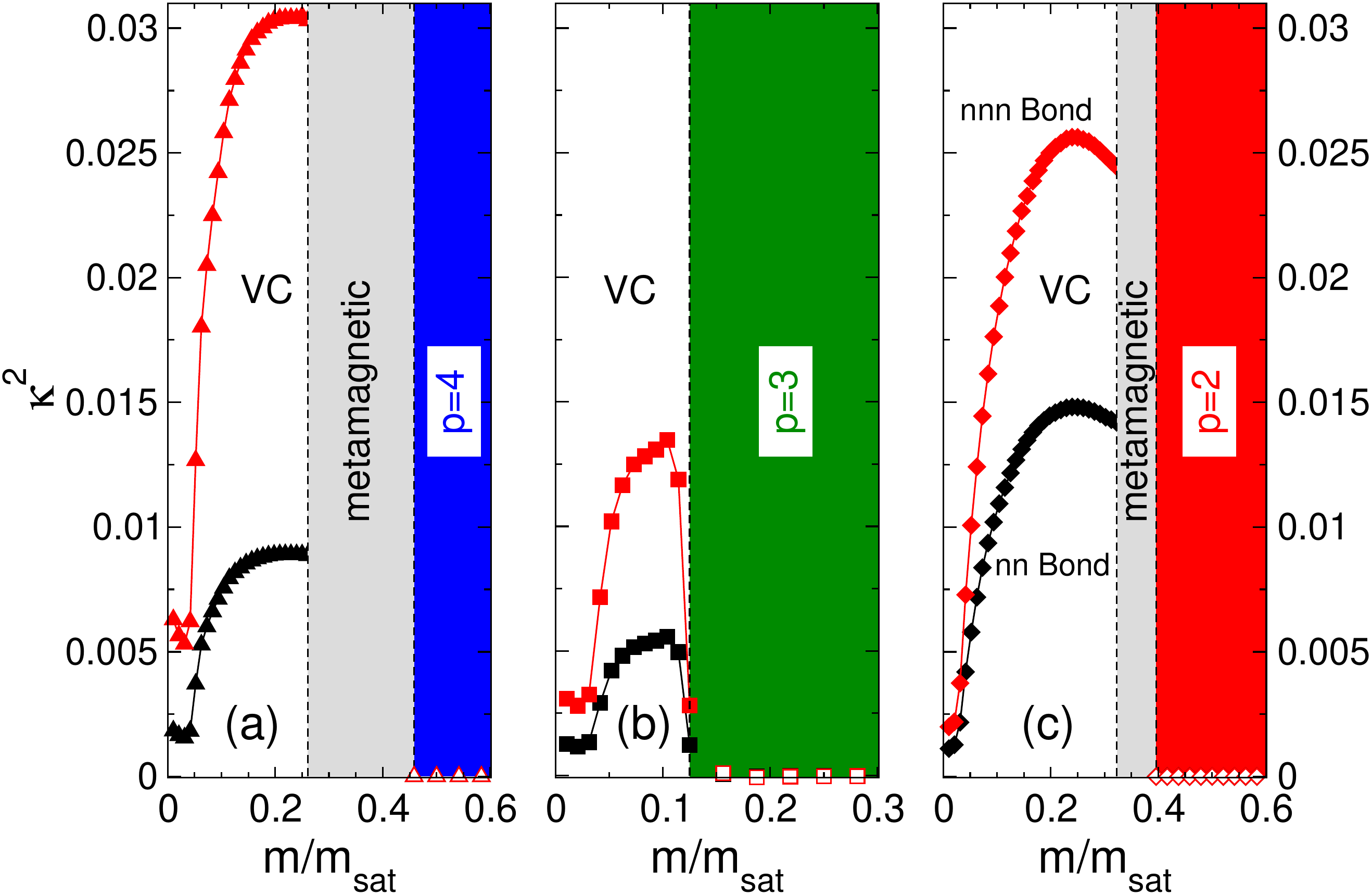}
\caption{(Color online) 
Squared vector chirality order parameter $\kappa^2$ (Eq.~\protect{\ref{eqn:kappasq}}) in the low magnetization phase 
for different values of $J_2/|J_1|$ as a function of $m/m_\mathrm{sat}$: (a) $J_2/|J_1|$=0.27, (b) $J_2/|J_1|$=0.32, (c) $J_2/|J_1|=0.38$. The order
parameter vanishes in the multipolar Luttinger liquids.
}
\label{fig:VCop}
\end{figure}
For $m>0$ we reveal a contiguous phase sustaining long range vector chiral order~\cite{Kolezhuk05} 
(breaking discrete parity symmetry), similar to phases recently discovered for $J_1,J_2>0$ \cite{Mcculloch08,Okunishi08}. 
Direct evidence for the presence of this phase is obtained from measurements of the squared vector chiral order parameter:
\begin{equation}
\kappa^2 (r,d):=\langle [\mathbf{S}_0\times \mathbf{S}_d]^z[\mathbf{S}_r\times \mathbf{S}_{r+d}]^z\rangle
\label{eqn:kappasq}
\end{equation}
In Fig.~\ref{fig:VCop} we display DMRG results for long distance correlations between $J_1$ bonds ($d=1$, black symbols) and  $J_2$ bonds 
($d=2$, red symbols) obtained on a $L=192$ system. The three chosen values of $J_2$ reflect positions underneath each of the three
spin-multipolar Luttinger liquids shown in Fig.~\ref{fig:PhaseDiagram}. 
The non-monotonic behavior of the correlations at very
small $m$ is probably a finite size artifact or convergence issue.
Beyond the long range order in the vector chirality,
the system behaves as a single channel Luttinger liquid (with central charge $c=1$, confirmed by our DMRG based entanglement entropy analysis~\cite{cft}) with critical incommensurate transverse spin correlation functions~\cite{Mcculloch08}.
The transition to the spin-multipolar phases at larger $m$ seems to occur generically via metamagnetic behavior, c.f.~left and right panels of Fig.~\ref{fig:VCop}. For the parameter set in the middle panel we expect the same behavior, but it can't be resolved
based on the system sizes used.

\paragraph{Multipolar Luttinger liquid phases} 

The Hamiltonian Eq.~(\ref{eqn:Hamiltonian}) presents unusual elementary step sizes $\Delta S^z>1$ in some extended $J_2/|J_1|$ and $m$ domains,
where $\Delta S^z$ is independent of the system size~\cite{Heidrichmeisner06}. 
This phenomenon has been explained based on the formation of bound states of $p=\Delta S^z$ magnons in the completely saturated state, and at finite $m/m_\mathrm{sat}$ a description in terms of a single component Luttinger liquid of bound states 
has been put forward~\cite{Vekua07,Kecke07}.
We have determined the extension of the $\Delta S^z=2,3,4$ regions in Fig.~\ref{fig:PhaseDiagram}, based on exact diagonalizations on systems sizes
up to 32 sites and DMRG simulations on systems up to 192 sites. The boundaries are in very good agreement with previous results~\cite{Heidrichmeisner06} where available. The $\Delta S^z=3$ and $4$ domains form lobes which are widest at $m_\mathrm{sat}$ and whose tips do not extend down to zero magnetization. The higher lobes are successively narrower in the $J_2$ direction. We have also searched for $\Delta S^z=5$ and higher regions, but found them to be unstable against a direct metamagnetic transition from the vector chiral phase to full saturation. Individual bound states of $p\geq5$ magnons do exist (see below), but they experience a too strong mutual attraction to be thermodynamically stable.

An exciting property of the Luttinger liquids of $p$ bound magnon states~\cite{Kecke07} is that the transverse spin correlations are exponentially decaying as a function of distance due to the binding,
while $p$-multipolar spin correlations
\begin{equation}
\langle \prod_{n=0}^{p-1} S^+_{0+n}
\prod_{n=0}^{p-1} S^-_{r+n}\rangle
\sim
(-1)^r\ \left(\frac{1}{r}\right) ^{1/K}
\label{eqn:multipolarcorrs}
\end{equation}
are critical with wave vector $\pi$ (multipolar correlations with $p'<p$ also decay exponentially). $p=2,3,4$ correspond to quadrupolar, octupolar and hexadecupolar correlations respectively.
Therefore they can be considered as one-dimensional analogues of spin multipolar ordered phases found in higher dimensions.
Another important correlation function is the longitudinal
spin correlator, which is also critical~\cite{Kecke07}:
\begin{equation}
\langle  S^z_{0}
S^z_{r}\rangle - m^2
\sim
\cos\left[\frac{(1- m/m_\mathrm{sat})\pi r}{p}\right]\ \left(\frac{1}{r}\right) ^{K}
\label{eqn:szcorrs}
\end{equation}
We verified numerically $c=1$ and determined the Luttinger parameter $K$ as a function of $m$ and $J_2$ by fits to the local $S^z$ profile in DMRG simulations.
An important information is contained in the crossover line $K = 1$ where $p$-multipolar and longitudinal spin correlations decay with the same 
exponent. This crossover line is shown for the three lobes in Fig.~\ref{fig:PhaseDiagram}. Close to saturation the spin-multipolar correlations 
dominate while towards the tip of the lobes the longitudinal spin correlations decay more slowly. The crossover line is rather flat in the
$J_2$ direction, but drops to lower $m$ values when going from $p=2$ to 3 and 4. This can be 
understood from the analogy with a Luttinger liquid of hardcore bosons of bound states which effectively becomes more dilute when increasing $p$ at constant $m$.
\begin{figure}[hbt]
\centering
\includegraphics[width=0.9\linewidth]{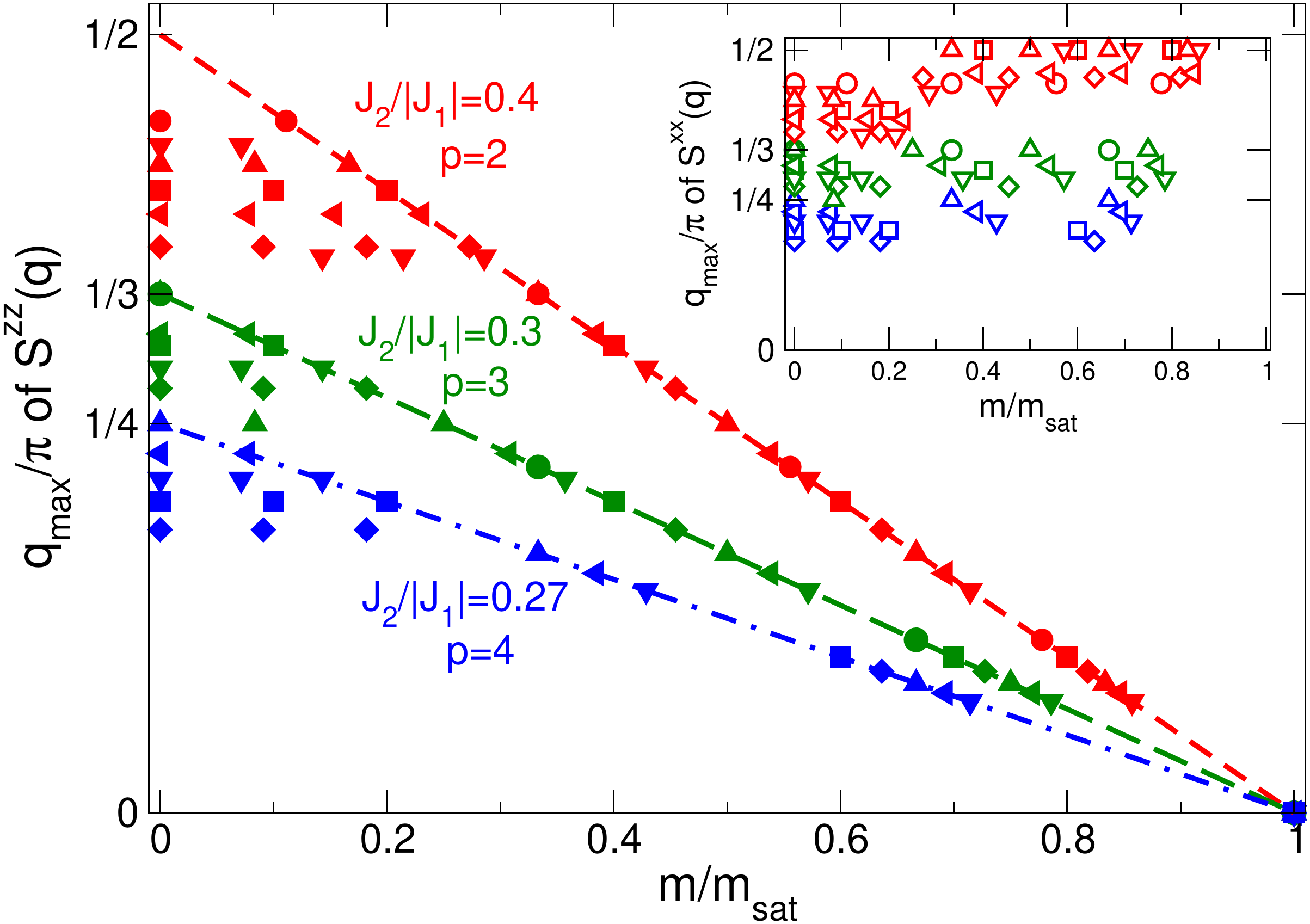}
\caption{(Color online) Wave vector $q$ maximizing the longitudinal (inset: transverse) spin structure factor as a function of the magnetization $m/m_\mathrm{sat}$ 
for selected $J_2/|J_1|$ parameters, each one belonging to a different $p$ sector at high $m$. 
The straight lines are guides to the eye and show $\pi(1-m/m_\mathrm{sat})/p$. 
The different symbols denotes ED results on system sizes ranging from 18 to 28 sites.}
\label{fig:StructFact}
\end{figure}

\paragraph{Interpretation}

After having established and characterized the phase diagram, we are lead to the intriguing
question whether there is a deeper relation between the spin multipolar phases which would enable us to 
understand their presence, and possibly predict related phenomena in other systems.

Let us first investigate how the longitudinal and transverse equal-time spin structure factors $\mathcal{S}^{zz}(q)$ 
and $\mathcal{S}^{xx}(q)$ evolve as a function of 
$m$. In Fig.~\ref{fig:StructFact} we display the location of the maximum
of  $\mathcal{S}^{zz}(q)$ and $\mathcal{S}^{xx}(q)$ (disregarding the $q=0$ peak in $\mathcal{S}^{zz}(q)$ due
to the total magnetization) for three representative $J_2$ values. At $m=0$ it is known that $\mathcal{S}(q)$ has a maximum
at an incommensurate position $q_\mathrm{max}(J_2)$, which is strongly renormalized compared to the classical expectation
$q^\mathrm{class}_\mathrm{max}(J_2)=\arccos(1/4J_2)$~\cite{Bursill}. In the low magnetization region, corresponding
to the vector chiral phase, the location of the maxima of both structure factors are only weakly dependent on $m$, and in a first 
approximation remain the same as for $m=0$.
However as $m$ is increased, the $q_\mathrm{max}$ of 
$\mathcal{S}^{zz}(q)$ locks onto a straight line with slope $-\pi/p$. 
It seems that if $q_\mathrm{max}(J_2)/\pi>1/3$ at $m=0$ then the magnetization process will enter the $p=2$ multipolar
phase at larger $m$. If instead $1/3>q_\mathrm{max}(J_2)/\pi>1/4$ at $m=0$ then the system will enter the $p=3$ phase.
Based on an extended analysis including several more $J_2/J_1$ values we are thus lead to conjecture that if 
\begin{equation}
1/p > q_\mathrm{max}(J_2)/\pi > 1/(p+1)\quad \mbox{ at } m=0,
\label{eq:locking_rule}
\end{equation}
then the higher $m$ region will lock onto the line with slope $-\pi/p$. According to the behavior of the longitudinal spin correlations in the multipolar Luttinger liquids 
(Eq.~\ref{eqn:szcorrs}), the multipolar liquid of order $p$ leads precisely to a slope of $-\pi/p$. 

A complementary point of view is provided by the consideration of a simple {\em classical} spiral.
The corresponding magnetic structure factor has a peak at the propagation vector $q$ of the spiral. 
It is straightforward to show that in such a state the in-plane multipolar 
correlations of order $p$ (analog to Eq.~\ref{eqn:multipolarcorrs}) have a propagation vector $q_p=p \cdot q$. 
Of course in a magnetically ordered state these multipolar correlations are not primary order parameters, but strong quantum 
fluctuations can wipe out the spin order, while a particular multipolar order parameter survives.
Applying this observation to the structure factor results, we can understand the rule~(\ref{eq:locking_rule}) as a locking of the multipolar-$p$
correlations to $\pi$, which precisely corresponds to the staggered nature of the multipolar-$p$ correlations in Eq.~\ref{eqn:multipolarcorrs}.

\begin{figure}[hbt]
\centering
\includegraphics[width=\linewidth]{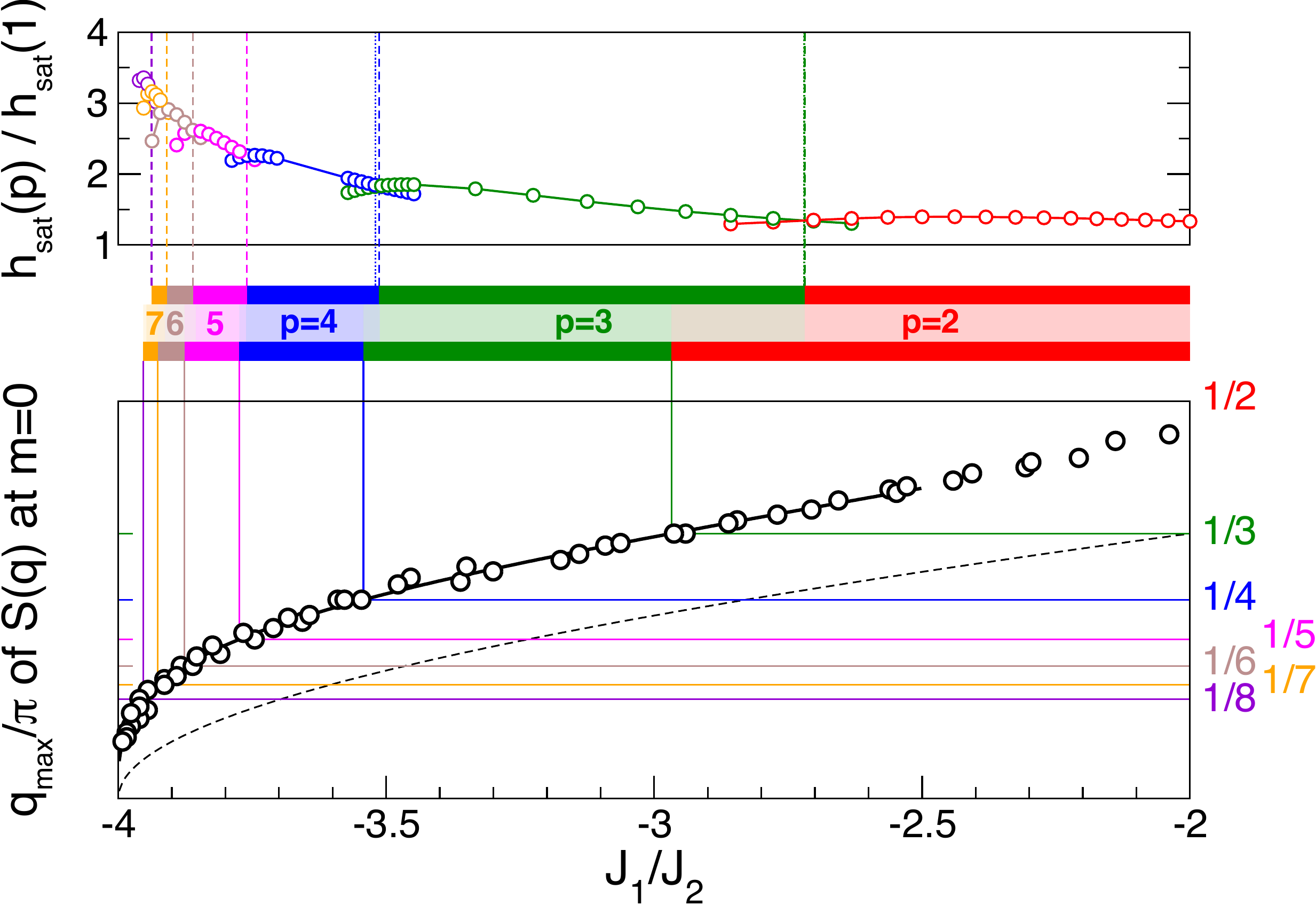}
\caption{
(Color online) Upper panel: ED results for the domain of stability of bound states of $p$ spin flips above the saturated
ferromagnetic state. Lower panel: Location $q_\mathrm{max}$ of the maximum of the zero field spin structure
factor $\mathcal{S}(q)$ as a function of $J_2$. The colored lines indicate the construction of the domain boundaries,
based on the locking rule Eq.~(\protect{\ref{eq:locking_rule}}). The thin dashed line displays $q^\mathrm{class}_\mathrm{max}$
for the classical model.}
\label{fig:InterpretedPhaseDiagram}
\end{figure}
Despite the fact that there are no stable multipolar Luttinger liquids with $p>4$, it is
possible to examine the process of individual bound state formation by pushing previous calculations from
$p\leq 4$ in Ref.~\cite{Kecke07} to $p=8$. We performed ED without truncation on system sizes
up to $L=64$, calculate the binding energy of $p$ flipped spins [expressed as $h_\mathrm{sat}(p)/h_\mathrm{sat}(p=1)$]
and display it in the upper panel of Fig.~\ref{fig:InterpretedPhaseDiagram}. We obtain a surprising series of stable bound states up to $p=8$,
which are successively smaller in their $J_2$ extent, suggesting that $J_2=1/4$ is an accumulation point of $p\rightarrow \infty$ bound states.
We now proceed to a comparison of the stability domain of the bound states to those predicted by the locking rule~(\ref{eq:locking_rule}) applied to the {\em zero field} structure factor. We fit a power law $q_\mathrm{max}(J_2) \sim  (J_2-1/4)^\gamma$ (with 
$\gamma\approx 0.29$) to the finite size transition points, as shown by the
bold black line in the lower panel of Fig.~\ref{fig:InterpretedPhaseDiagram}. This fitted line is used to construct the boundaries,
giving rise to the lower color bar. 
The agreement between the transition boundaries in the upper and lower panels is excellent (apart from the $p=2$ to $p=3$ transition).
It is striking that we are able to reproduce the domain of stability of bound states of magnons at the saturation magnetization,
based solely on the spin structure factor obtained at zero field. This constitutes strong evidence for an approximate validity of the 
locking rule Eq.~\ref{eq:locking_rule} and the presence of a locking mechanism which pins the multipolar correlations to $\pi$ at
larger $m$ as opposed to incommensurate transverse spin correlations for lower $m$.
The transition from $p=2$ to $p=3$ is somewhat shifted compared to the prediction, and we attribute this discrepancy to
the formation of a peculiar incommensurate $p=2$ bound state~\cite{Chubukov91,Kecke07}, leading to a collapse of the kinetic
energy of the $p=2$ bound state. 

Our considerations presented above are not restricted to one dimension. Indeed the recently reported quadrupolar~\cite{shannon06} 
and octupolar phases~\cite{momoi06} in frustrated $S=1/2$ ferromagnets in a field also allow an interpretation as fluctuation destabilized 
magnetic states. In the $J_1$-$J_2$ model on the square lattice there is a magnetically ordered striped phase [$Q=(\pi,0)$] in the 
neighborhood of the "bond nematic" phase. The nearest neighbor bond quadrupolar correlation functions in such a 
classical striped state have momentum $(0,0)$ and are odd under a $\pi/2$ rotation of one bond. Interestingly this is exactly the 
signature of the correlation function reported in the magnetically disordered, "bond nematic" phase. 
Similarly for the "triatic" or octupolar phase, there is a canted magnetic three-sublattice state $[Q=(4\pi/3,0)]$ 
neighboring the "triatic" phase at high fields~\cite{momoi06}. The derivation of the octupolar correlations on a triangle in the classical state again yields 
octupolar correlations matching the symmetry of the bound states reported in the "triatic" phase.
Based on the success of our considerations we speculate that it could be possible to stabilize a uniform hexadecupolar phase in the
square lattice $J_1-J_3$ model with ferromagnetic nearest neighbor interactions.

\paragraph{Experimental relevance}
The different phases we found should be detectable experimentally in the quasi-1D compounds 
studied in Refs.~\cite{Hase04,Enderle05, Drechsler07}.
In neutron scattering experiments these phases are expected to manifest as follows: i)~a Bragg peak at low magnetizations corresponding
to spiral order in the plane transverse to the magnetic field, ii)~a Bragg peak at intermediate magnetizations at wave vector $\pi (1-m/m_\mathrm{sat})/p$ corresponding to SDW(p) magnetic order modulated along the field and iii)~the absence of nontrivial magnetic Bragg peaks at high fields in the spin multipolar ordered phases.

\paragraph{Conclusions}

We have established the phase diagram of the frustrated ferromagnetic $S=1/2$ Heisenberg chain in a uniform magnetic field.
We provided an explanation for the appearance of a large number of bound states upon approaching the Lifshitz point at
$J_2=1/4$, based on a locking mechanism leading to spin multipolar phases. It is an open problem whether these fluctuation 
driven multipolar phases and the locking mechanism also appear for $S>1/2$. In a recent paper by Hikihara {\it et al.}~\cite{hikihara}
a phase diagram very similar to ours has been obtained.

\acknowledgments
We thank A. Honecker, C.L.~Henley and A. Kolezhuk for interesting discussions. 
We acknowledge support by the Swiss National Fund. The computations were performed at the CSCS (Manno, Switzerland).


\begin{thebibliography}{99}
\bibitem{OldSpirals}
T.A. Kaplan, Phys. Rev. {\bf 116}, 888 (1959);
J. Villain, J. Phys. Chem. Solids {\bf 11}, 303 (1959);
A. Yoshimori, J. Phys. Soc. Jpn. {\bf 14}, 807 (1959).
\bibitem{Villain}
J. Villain,
J. Phys. C: Solid State Phys. {\bf 10}, 4793 (1977).
\bibitem{Kawamura}
H. Kawamura,
J. Phys.: Condens. Matter {\bf 10}, 4707 (1998).
\bibitem{Chandra91}
P. Chandra and P. Coleman,
Phys. Rev. Lett. {\bf 66}, 100 (1991).
\bibitem{Andreev84}
  A.F. Andreev and I.A. Grishchuk,
  Sov. Phys. JETP {\bf 60}, 267 (1984).
\bibitem{Laeuchli05} 
A. L\"auchli {\it et al.},
Phys. Rev. Lett. {\bf 95}, 137206 (2005).
\bibitem{Cinti08}
F.~Cinti {\it et al.},
Phys. Rev. Lett. {\bf 100}, 057203 (2008).
\bibitem{Seki08}
S.~Seki {\it et al.}, 
Phys. Rev. Lett. {\bf 100}, 127201 (2008).
\bibitem{shannon06}
N. Shannon {\it et al.},
Phys. Rev. Lett. \textbf{96}, 027213 (2006).
\bibitem{momoi06} 
T. Momoi {\it et al.},
Phys. Rev. Lett. \textbf{97}, 257204 (2006).
\bibitem{White92}
S.R.~White,
Phys. Rev. Lett. {\bf 69}, 2863 (1992).
\bibitem{ItoiQin01}
C. Itoi and S. Qin,
Phys. Rev. B {\bf 63}, 224423 (2001).
\bibitem{Cabra00}
D.C.~Cabra {\it et al.},
Eur. Phys. Jour. B {\bf  13}, 55 (2000).
\bibitem{Hamada88} 
T. Hamada {\it et al.},
J. Phys. Soc. Jpn. \textbf{57}, 1891 (1988).
\bibitem{Heidrichmeisner06} 
F. Heidrich-Meisner {\it et al.},
Phys. Rev. B {\bf 74}, 020403(R) (2006).
\bibitem{Vekua07} 
T.~Vekua {\it et al.},
Phys. Rev. B \textbf{76}, 174420 (2007).
\bibitem{Kecke07} 
L.~Kecke {\it et al.},
Phys. Rev. B {\bf 76}, 060407(R) (2007).
\bibitem{Hase04} 
M. Hase \textit{et al.}, Phys. Rev. B \textbf{70}, 104426 (2004).
\bibitem{Enderle05} 
M. Enderle \textit{et al.}, 
Europhys. Lett. \textbf{70}, 237 (2005).
\bibitem{Drechsler07}
S.-L. Drechsler \textit{et al.},
Phys. Rev. Lett. {\bf 98}, 077202 (2007).
\bibitem{Kolezhuk05} 
A. Kolezhuk and T. Vekua, 
Phys. Rev. B \textbf{72}, 094424 (2005).
\bibitem{Mcculloch08} 
I.P. McCulloch \textit{et al.}, 
Phys. Rev. B {\bf 77}, 094404 (2008).
\bibitem{Okunishi08}
K.~Okunishi, 
J. Phys. Soc. Jpn. {\bf 77}, 114004 (2008).
\bibitem{cft}
P. Calabrese and J. Cardy,
J. Stat. Mech. (2004) P06002.
\bibitem{Bursill}
R. Bursill {\it et al.},
J. Phys. Condens. Matter {\bf 7}, 8605 (1995).
\bibitem{Chubukov91}
A.V. Chubukov,
Phys. Rev. B {\bf 44}, 4693 (1991).
\bibitem{hikihara}
T. Hikihara {\it et al.},
Phys. Rev. B {\bf 78}, 144404 (2008).
\end{thebibliography}
\end{document}